\documentclass[%
reprint,
aps, prapplied,
 amsmath,amssymb,
 longbibliography
]{revtex4-1}

\usepackage[colorlinks=true,linkcolor=blue]{hyperref}%
\expandafter\ifx\csname package@font\endcsname\relax\else
 \expandafter\expandafter
 \expandafter\usepackage
 \expandafter\expandafter
 \expandafter{\csname package@font\endcsname}%
\fi
\usepackage[normalem]{ulem}
\usepackage{graphicx}
\usepackage{dcolumn}
\usepackage{bm}
\usepackage{epstopdf}
\usepackage{epsfig}
\graphicspath{ {Plots/} }

\begin{document}


\title{Band gap and band offset of Ga$_2$O$_3$ and (Al$_x$Ga$_{1-x}$)$_2$O$_3$ alloys}

\author{Tianshi Wang}
\affiliation{Department of Materials Science and Engineering, University of Delaware, Newark, Delaware 19716, USA}

\author{Wei Li}
\affiliation{Department of Materials Science and Engineering, University of Delaware, Newark, Delaware 19716, USA}

\author{Chaoying Ni}
   \email{cni@udel.edu}
\affiliation{Department of Materials Science and Engineering, University of Delaware, Newark, Delaware 19716, USA}

\author{Anderson Janotti}
  \email{janotti@udel.edu}
\affiliation{Department of Materials Science and Engineering, University of Delaware, Newark, Delaware 19716, USA}
%


\begin{abstract}

Ga$_2$O$_3$ and (Al$_x$Ga$_{1-x}$)$_2$O$_3$ alloys are promising materials for solar-blind UV photodetectors and high-power transistors.  Basic key parameters in the device design, such as band gap variation with alloy composition and band offset between Ga$_2$O$_3$ and (Al$_x$Ga$_{1-x}$)$_2$O$_3$,  are yet to be established.
 Using density functional theory with the HSE hybrid functional, we compute formation enthalpies, band gaps, and band edge positions of  (Al$_x$Ga$_{1-x}$)$_2$O$_3$ alloys in the  monoclinic ($\beta$)  and corundum ($\alpha$)  phases.  We find the formation enthlapies of (Al$_x$Ga$_{1-x}$)$_2$O$_3$ alloys are significantly lower than of (In$_x$Ga$_{1-x}$)$_2$O$_3$, and that (Al$_x$Ga$_{1-x}$)$_2$O$_3$ with $x$=0.5 can be considered as an ordered compound AlGaO$_3$ in the monoclinic phase, with Al occupying the octahedral sites and Ga occupying the tetrahedral sites.
 The band gaps of the alloys range from  4.69 to 7.03 eV for $\beta$-(Al$_x$Ga$_{1-x}$)$_2$O$_3$ and from 5.26 to 8.56 eV for $\alpha$-(Al$_x$Ga$_{1-x}$)$_2$O$_3$.  Most of the band offset of the (Al$_x$Ga$_{1-x}$)$_2$O$_3$ alloy  arises from the discontinuity in the conduction band. 
Our results are used to explain the available experimental data, and consequences for designing modulation-doped field effect transistors (MODFETs) based on (Al$_x$Ga$_{1-x}$)$_2$O$_3$/Ga$_2$O$_3$ are discussed. 

\end{abstract}

\maketitle

Ga$_2$O$_3$ has been intensively investigated as a wide-band-gap semiconductor for high-power electronics \cite{Higashiwaki2012, Masataka2016a,   Jessen2017} and UV solar-blind phtodetectors \cite{Oshima2008, Oshima2007}. It is available as large single crystals \cite{Aida2008} suitable for high-quality epitaxial thin-film growth by metalorganic chemical vapor deposition (MOCVD) \cite{Du2015, Baldini2017} and molecular beam epitaxy (MBE) \cite{Villora2006, Oshima2007}; It displays high breakdown electric field \cite{Higashiwaki2012}, and the Baliga figure of merit exceeds that of SiC and GaN \cite{Jessen2017}; It can be easily doped $n$-type, and band gap engineering can be accomplished by incorporating In and Al, adding great flexibility to device design.  Modulation doping of (Al$_x$Ga$_{1-x}$)$_2$O$_3$/Ga$_2$O$_3$ heterostructures can be used to separate the ionized donors in the (Al$_x$Ga$_{1-x}$)$_2$O$_3$ layer from the conduction electrons in the Ga$_2$O$_3$ layer \cite{Krishnamoorthy2017, Oshima2017, Ahmadi2017a,Kang2017}, providing a boost to the electron mobility to about 500 cm$^2$V$^{-1}$s$^{-1}$ \cite{Ma2016,Zhang2018,Krishnamoorthy2017} by suppressing scattering from the ionized impurities. 
Simulated band diagrams and two-dimension electron gas (2DEG) profile of MODFETs based on (Al$_x$Ga$_{1-x}$)$_2$O$_3$/Ga$_2$O$_3$ assumed that the discontinuity in the band offset appears solely on the conduction band \cite{Krishnamoorthy2017}.  However this assumption has not been based on firm experimental evidence or first-principles calculations.
The band gap of the (Al$_x$Ga$_{1-x}$)$_2$O$_3$ alloy and the band offset between the 
 (Al$_x$Ga$_{1-x}$)$_2$O$_3$ and Ga$_2$O$_3$ are key parameters in the device design and are yet to be established.

\begin{figure}
\includegraphics[width=8.6cm]{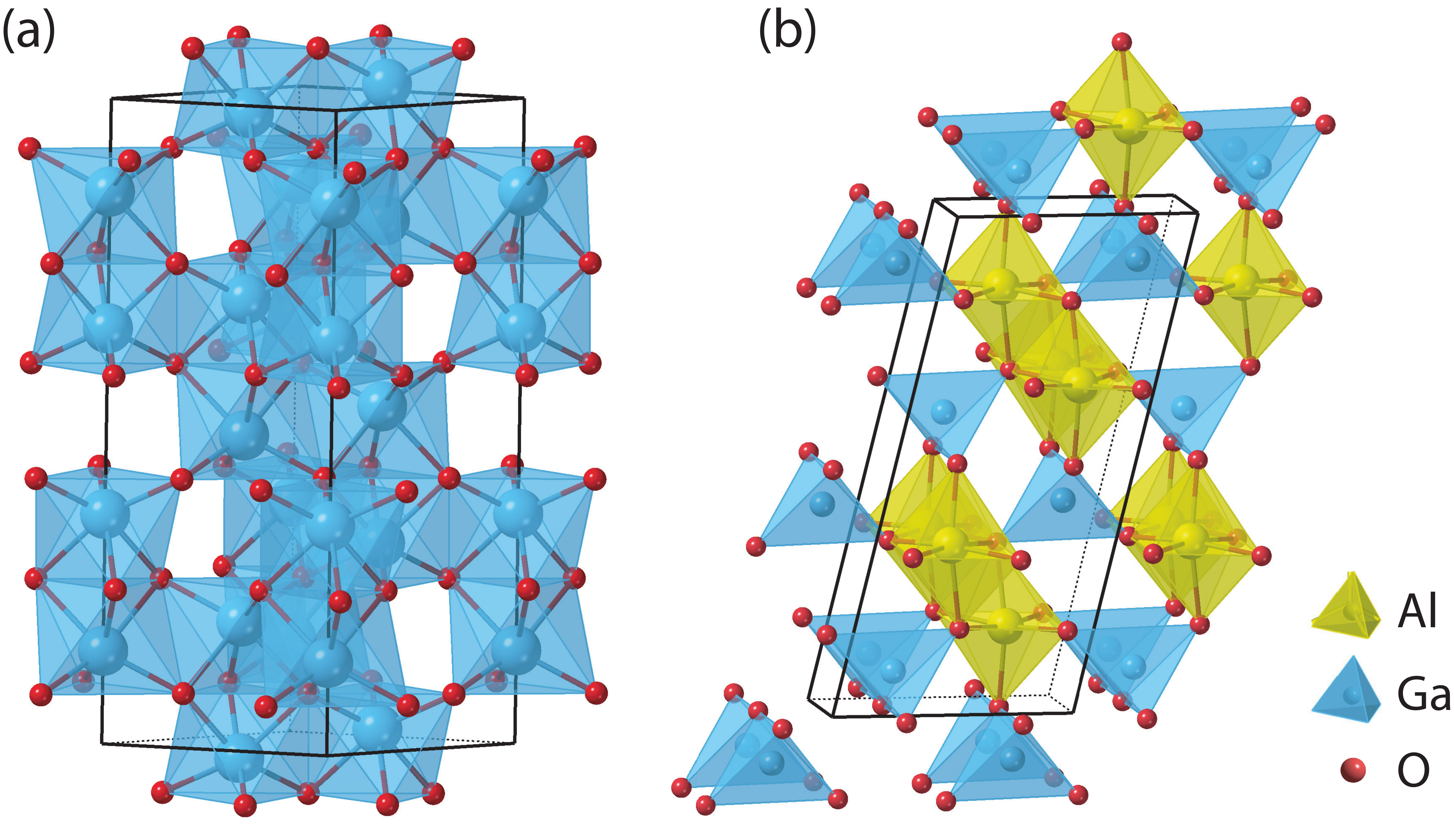}
\caption{Conventional unit cells of (a) $\alpha$-Ga$_2$O$_3$ and (b) $\beta$-AlGaO$_3$ compound. In the latter, the Al atoms occupy octahedral sites and Ga occupy tetrahedral sites, maintaining the same structure as its parent compounds $\beta$-Ga$_2$O$_3$ and $\theta$-Al$_2$O$_3$. }
\label{Fig1}
\end{figure}

Ga$_2$O$_3$ and (Al$_x$Ga$_{1-x}$)$_2$O$_3$ can be made in the monoclinic ($\beta$) and in the corundum ($\alpha$) phase, as shown in Fig.~\ref{Fig1}. Bulk and thin films of $\beta$-(Al$_x$Ga$_{1-x}$)$_2$O$_3$ have been obtained using solution combustion synthesis \cite{Krueger2016}, pulsed laser deposition (PLD) \cite{Zhang2014b}, and oxygen plasma-assisted molecular beam epitaxy (MBE) \cite{Krishnamoorthy2017, Oshima2017, Ahmadi2017a}, while $\alpha$-(Al$_x$Ga$_{1-x}$)$_2$O$_3$ have been grown on sapphire substrates by chemical vapor deposition (CVD) for Al content of up to 81\% \cite{Ito2012, Uchida2016}.
Band gaps of (Al$_x$Ga$_{1-x}$)$_2$O$_3$ for selected Al content have been reported \cite{Ito2012, Uchida2016, Krueger2016, Zhang2014b, Feng2017}, but band offsets between Ga$_2$O$_3$ and (Al$_x$Ga$_{1-x}$)$_2$O$_3$,  which are much more challenging to obtain experimentally, are still unknown. 

Using density functional theory and the HSE hybrid functional, we investigate the formability of (Al$_x$Ga$_{1-x}$)$_2$O$_3$ alloys in both $\beta$ and $\alpha$ phases, their band gaps as a function of Al concentration, and the band offsets of the (Al$_x$Ga$_{1-x}$)$_2$O$_3$ alloys with respect to Ga$_2$O$_3$ and Al$_2$O$_3$. 
In the following, we first describe the details of the calculations, present the results for structural parameters and stability of the alloys, and then discuss the results for band gaps and band offsets for both $\beta$ and $\alpha$ phases and the implications for device design.

\begin{table*}
\setlength\tabcolsep{8pt} 
\caption{\label{table1}
Calculated lattice parameters, formation enthalpy ($\Delta H$), and band gaps  ($E_g$) of  Al$_2$O$_3$ and  Ga$_2$O$_3$ in corundum  ($\alpha$-Al$_2$O$_3$ and $\alpha$-Ga$_2$O$_3$) and monoclinic  ($\theta$-Al$_2$O$_3$ and $\beta$-Ga$_2$O$_3$) structures. The indirected and direct band gaps are denoted as (i) and (d), respectively. 
Note that monoclinic  Al$_2$O$_3$ is often denoted as the $\theta$ phase in the literature, i.e., $\theta$- Al$_2$O$_3$ shares the same crystal structure as $\beta$-Ga$_2$O$_3$ \cite{Yamaguchi1970, Zhou1991, husson1996structural}.
}
\begin{ruledtabular}
\begin{tabular}{lcccccccc}
\rule{0pt}{10pt} & \multicolumn {2}{c}{$\alpha$-Al$_2$O$_3$}& \multicolumn {2}{c}{ $\theta$-Al$_2$O$_3$} &  \multicolumn {2}{c}{ $\alpha$-Ga$_2$O$_3$}&   \multicolumn {2}{c}{$\beta$-Ga$_2$O$_3$}\\

\rule{0pt}{10pt}  & Calc.& Expt. & Calc.& Expt. & Calc.& Expt. & Calc.& Expt. \\
\hline
$a$(\AA)& 4.774 & 4.76 \footnotemark[1] & 11.808  &11.795 \footnotemark[2]  & 5.005 &	4.983 \footnotemark[3] &	12.276&12.214 \footnotemark[4] \\ 
$b$(\AA)&  &   &2.921 &2.910 \footnotemark[2] &	  &	 & 3.050 &  3.037 \footnotemark[4] \\ 
$c$(\AA) & 13.013 & 12.99 \footnotemark[1]  & 5.636 & 5.621 \footnotemark[2] & 13.454 & 13.433 \footnotemark[3]&5.811 & 5.798 \footnotemark[4] \\
$\beta$(deg)&  & & 104.08 &103.79 \footnotemark[2]  &  & & 103.72& 103.83 \footnotemark[4] \\
$\Delta H$ (eV/f.u.)& -15.753 &-16.971 \footnotemark[5]  & -15.561 &   &-9.824  &  &-9.870 & -11.30 \footnotemark[6] \\  
$E_\textrm{g}$ (eV) & 8.56 (d) & 8.8 \footnotemark[7]    & 7.03 (i) &  &5.26 (i) &5.3 \footnotemark[8]    & 4.69 (i)& 4.48 \footnotemark[9], 4.9 \footnotemark[10]
\end{tabular}
\begin{minipage}{\textwidth}
     \begin{flushleft}
\textsuperscript{a} Ref. \onlinecite{newnham1962refinement}; \textsuperscript{b} Ref. \onlinecite{husson1996structural}; \textsuperscript{c} Ref. \onlinecite{doi:10.1063/1.1840945}; \textsuperscript{d} Ref. \onlinecite{aahman1996reinvestigation}; \textsuperscript{e} Ref. \onlinecite{ghosh1977standard}; \textsuperscript{f} Ref. \onlinecite{Barin2008}; \textsuperscript{g} Ref. \onlinecite{French1990}; \textsuperscript{h} Ref. \onlinecite{Shinohara2008}; \textsuperscript{i} Ref. \onlinecite{onuma2015valence}; \textsuperscript{j} Ref. \onlinecite{Orita2000}.
      \end{flushleft}
\end{minipage}

\end{ruledtabular}
\end{table*}

The calculations are based on the density functional theory (DFT) \cite{PhysRev.136.B864, Kohn1965} with the projector augmented-wave method (PAW) \cite{Blochl1994} as implemented in the VASP code \cite{Kresse1993, Kresse1994}. The $d$ states of Ga are included as valence states, and a plane-wave cutoff energy of 400 eV is employed. We use Perdew-Burke-Ernzerhof functional revised for solids (PBEsol) \cite{perdew2008restoring}  to relax all structures. To overcome the severe underestimation of band gaps  in the DFT-PBEsol functional, we employed the screened hybrid functional of Heyd, Scuseria, and Ernzerhof (HSE06) \cite{Heyd2003,HSE06}. In the HSE hybrid functional, the nonlocal Hartree-Fock exchange is mixed with the generalized gradient approximation (GGA) \cite{Perdew1996} exchange  in the short-range part. The mixing parameter in HSE is fixed to 32\% for all the calculations. We find that this choice of mixing parameter gives band gaps of the parent compounds Ga$_2$O$_3$ and Al$_2$O$_3$ in good agreement with experimental values. Note that, conventionally, monoclinic  Al$_2$O$_3$ is often denoted as the $\theta$ phase in the literature, i.e., $\theta$-Al$_2$O$_3$ shares the same crystal structure as $\beta$-Ga$_2$O$_3$ \cite{Yamaguchi1970, Zhou1991, husson1996structural}.
The calculated lattice parameters, formation enthalpies, and band gaps for Al$_2$O$_3$ and Ga$_2$O$_3$ are listed in  Table~\ref{table1} along with the available experimental data. The $\theta$-Al$_2$O$_3$, $\alpha$-Ga$_2$O$_3$, and $\beta$-Ga$_2$O$_3$ have indirect gaps where the valence-band maximum is higher than the valence-band edge at $\Gamma$ by 0.16, 0.24, and 0.04 eV, respectively, as previously reported \cite{Peelaers2015a}.

We simulate  (Al$_x$Ga$_{1-x}$)$_2$O$_3$  random alloys  using special quasirandom structures (SQS) \cite{Zunger1990} generated using the {\em mcsqs} code of the Alloy Theoretic Automated Toolkit (ATAT) \cite{vandeWalle2013}.  This method can generate optimal periodic supercells comparable to true disordered structures based on a Monte Carlo simulated annealing loop with an objective function that seeks to perfectly match the maximum number of correlation functions \cite{vandeWalle2013}.  

We use supercells containing 80 atoms for the $\alpha$ and 120 atoms for the $\beta$ phase. 
In the case of $\alpha$-Ga$_2$O$_3$, all Ga sites are equivalent; therefore, Al tends to replace Ga randomly. However, in the case of $\beta$-Ga$_2$O$_3$, half of Ga atoms are at octahedral sites and the other half at tetrahedral sites. We find that Al strongly prefers octahedral sites, i.e., Al sitting at octahedral sites is $\sim$0.2 eV per Al atom lower in energy than Al sitting at tetrahedral sites.  Therefore, in the generation of SQS structures to simulate $\beta$-(Al$_x$Ga$_{1-x}$)$_2$O$_3$ alloys, we assumed that Al  atoms occupy only octahedral sites for $x \leq 0.5$. For $x >0.5$, Al atoms exceeding $x>0.5$ randomly replace Ga at tetrahedral sites since all octahedral sites are already filled.

\begin{figure}
\includegraphics[width=7cm]{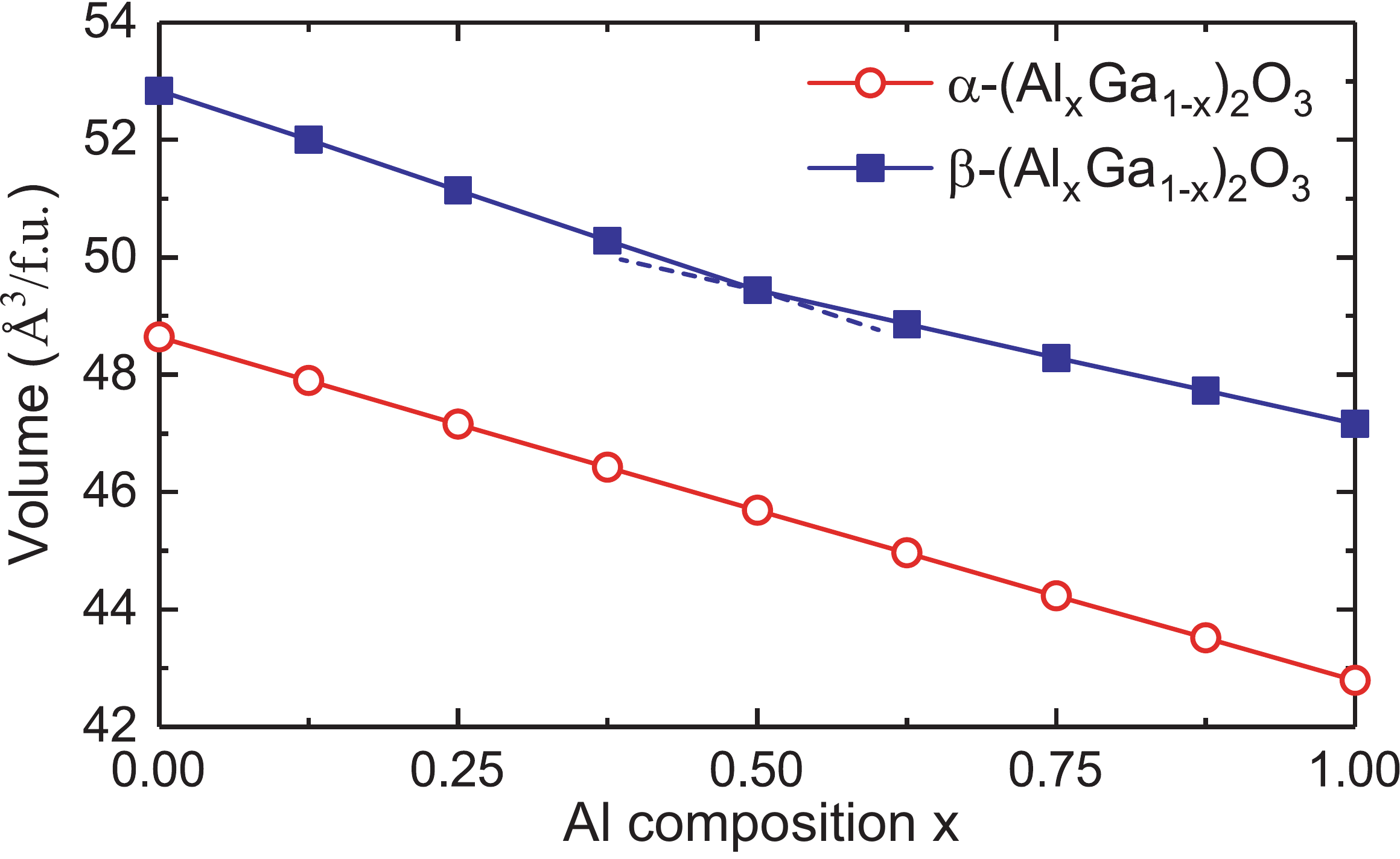}
\caption{Calculated equilibrium volume of $\alpha$ and  $\beta$-(Al$_x$Ga$_{1-x}$)$_2$O$_3$ alloys as a function of Al concentration ($x$). }
\label{Fig2}
\end{figure}

Figure~\ref{Fig2} shows the calculated equilibrium volume of (Al$_x$Ga$_{1-x}$)$_2$O$_3$ alloys as a function of Al fraction. 
For $\alpha$-(Al$_x$Ga$_{1-x}$)$_2$O$_3$, the volume varies linearly with Al composition, following Vegard's law.
For $\beta$ -(Al$_x$Ga$_{1-x}$)$_2$O$_3$, the volume also decreases monotonically with Al fraction, but exhibit a change in slope at $x = 0.5$. This discontinuity in the slope 
is attributed to Al occupying the tetrahedral sites for $x >0.5$.  This trend was recently observed by Krueger {\em et al.} \cite{Krueger2016}

\begin{figure}
\includegraphics[width=8.6cm]{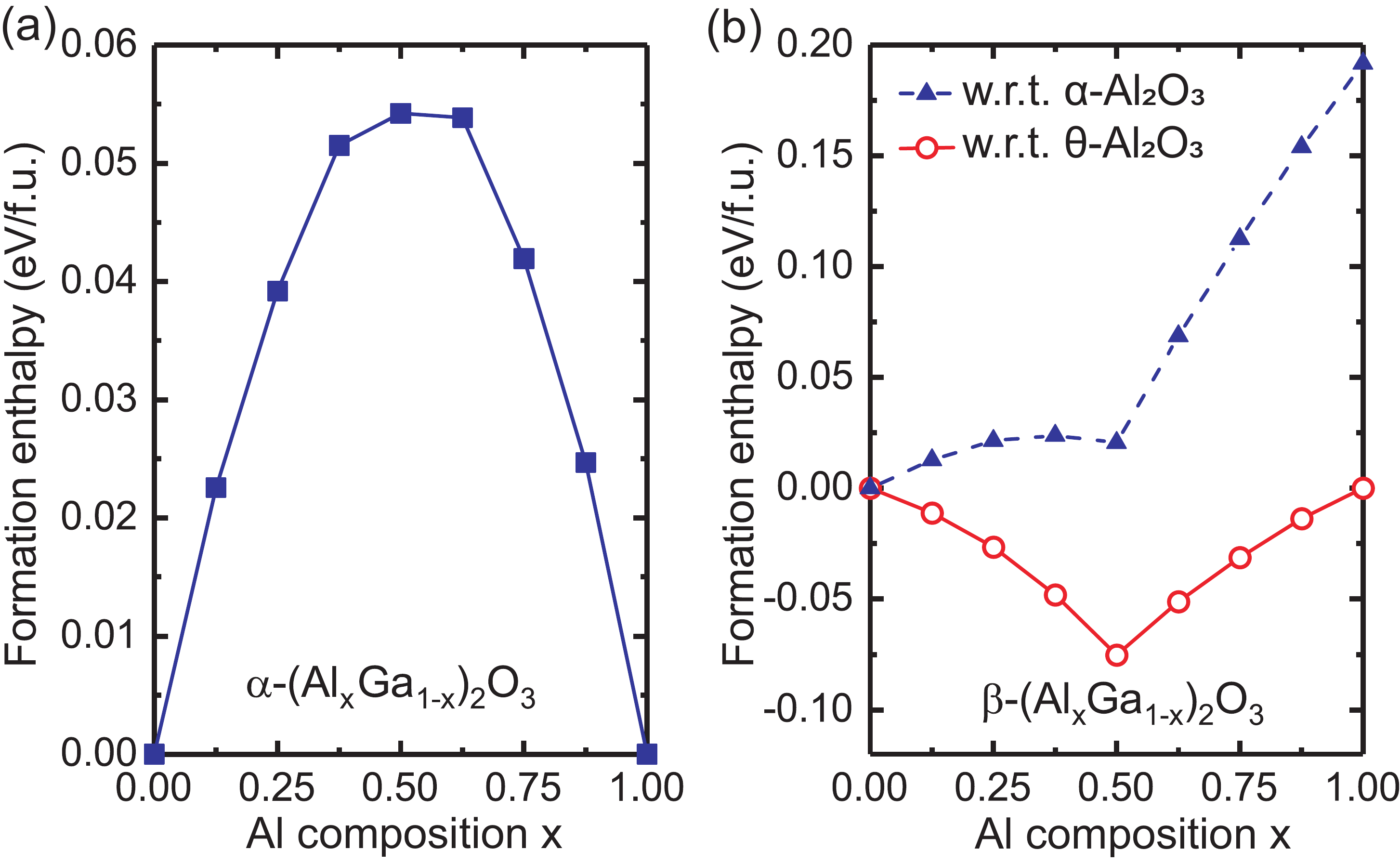}
\caption{Formation enthalpies of (a)  $\alpha$ and  (b) $\beta$-(Al$_x$Ga$_{1-x}$)$_2$O$_3$ with respect to the parent compounds Ga$_2$O$_3$ and Al$_2$O$_3$. For $\beta$-(Al$_x$Ga$_{1-x}$)$_2$O$_3$ , formation enthalpies calculated with respect to (w.r.t.) $\beta$-$\text{Ga}_2\text{O}_3$+$\theta$-$\text{Al}_2\text{O}_3$ (red circles) and  $\beta$-$\text{Ga}_2\text{O}_3$+$\alpha$-$\text{Al}_2\text{O}_3$ (blue triangles) are shown. }
\label{Fig3}
\end{figure}

The calculated formation enthalpies of Ga$_2$O$_3$  and Al$_2$O$_3$, defined as the total energy of the compound minus the total energies of the elemental bulk phases, are listed
 in Table~\ref{table1}.  We find that the formation enthalpy of $\alpha$-Ga$_2$O$_3$ is only 46 meV/f.u. higher than $\beta$-Ga$_2$O$_3$.  In contrast, $\alpha$-Al$_2$O$_3$ is much more stable than $\theta$-Al$_2$O$_3$, by 192 meV/f.u.  This is again attributed to the preference of Al occupying octahedral sites, by $\sim0.2$ eV per cation.
Figure~\ref{Fig3}(a) and (b) show the formation enthalpies of $\alpha$ and $\beta$-(Al$_x$Ga$_{1-x}$)$_2$O$_3$, defined as:

\begin{equation} \label{eq1}
\begin{split}
\Delta H[(\text{Al}_{x}\text{Ga}_{1-x})_2\text{O}_3] = &E[(\text{Al}_{x}\text{Ga}_{1-x})_2\text{O}_3] -  x E[\text{Al}_2\text{O}_3]\\
 &- (1-x) E[\text{Ga}_2\text{O}_3]  \,,
\end{split}
 \end{equation}
 where $E[(\text{Al}_{x}\text{Ga}_{1-x})_2\text{O}_3]$ is the total energy of the SQS supercell structure representing the random alloy, and $E[\text{Ga}_2\text{O}_3]$ and  $E[\text{Al}_2\text{O}_3]$ are the total energies of the parent compounds Ga$_2$O$_3$ and Al$_2$O$_3$ in the same supercell.
 The formation enthalpies of $\alpha$-(Al$_x$Ga$_{1-x}$)$_2$O$_3$ are relatively small compared to other alloys. For example, at $x=0.5$, the formation enthalpy of $\alpha$-(In$_x$Ga$_{1-x}$)$_2$O$_3$ is $\sim$300 meV/f.u.\cite{Peelaers2015}, compared to 55 meV/f.u. for (Al$_x$Ga$_{1-x}$)$_2$O$_3$. This indicates that $\alpha$-(Al$_x$Ga$_{1-x}$)$_2$O$_3$ alloys are more likely to form at all Al compositions.  
In the case of $\beta$-(Al$_x$Ga$_{1-x}$)$_2$O$_3$, we find a stable ordered compound AlGaO$_3$ for 50\% Al content. 
If taken with respect to $\alpha$-$\text{Al}_2\text{O}_3$ (dashed line), which is the most stable phase of $\text{Al}_2\text{O}_3$, the formation enthalpy of $\beta$-(Al$_x$Ga$_{1-x}$)$_2$O$_3$ rapidly increases with Al composition above 50\%.  Again, this is explained by the fact that Al strongly prefer to occupy the octahedral sites for $x \leq 0.5$, but endup occupying the only available tetrahedral sites for $x > 0.5$.

Therefore, for Al concentrations approaching 100\%,  we predict that (Al$_x$Ga$_{1-x}$)$_2$O$_3$ alloys strongly favor the corundum or $\alpha$ phase.  This explains why single monoclinic phase at $0 \leq x < 0.8$ and mixed corundum and monoclinic phases for $0.8 \leq x < 1$  have been observed by solution combustion synthesis \cite{Krueger2016}.  
 We note, however, that thin films of $\beta$-(Al$_x$Ga$_{1-x}$)$_2$O$_3$ alloys with Al content up to 96\% have been reported using pulsed laser deposition (PLD) \cite{Zhang2014b}.

\begin{figure}[b!]
\includegraphics[width= 7.5cm]{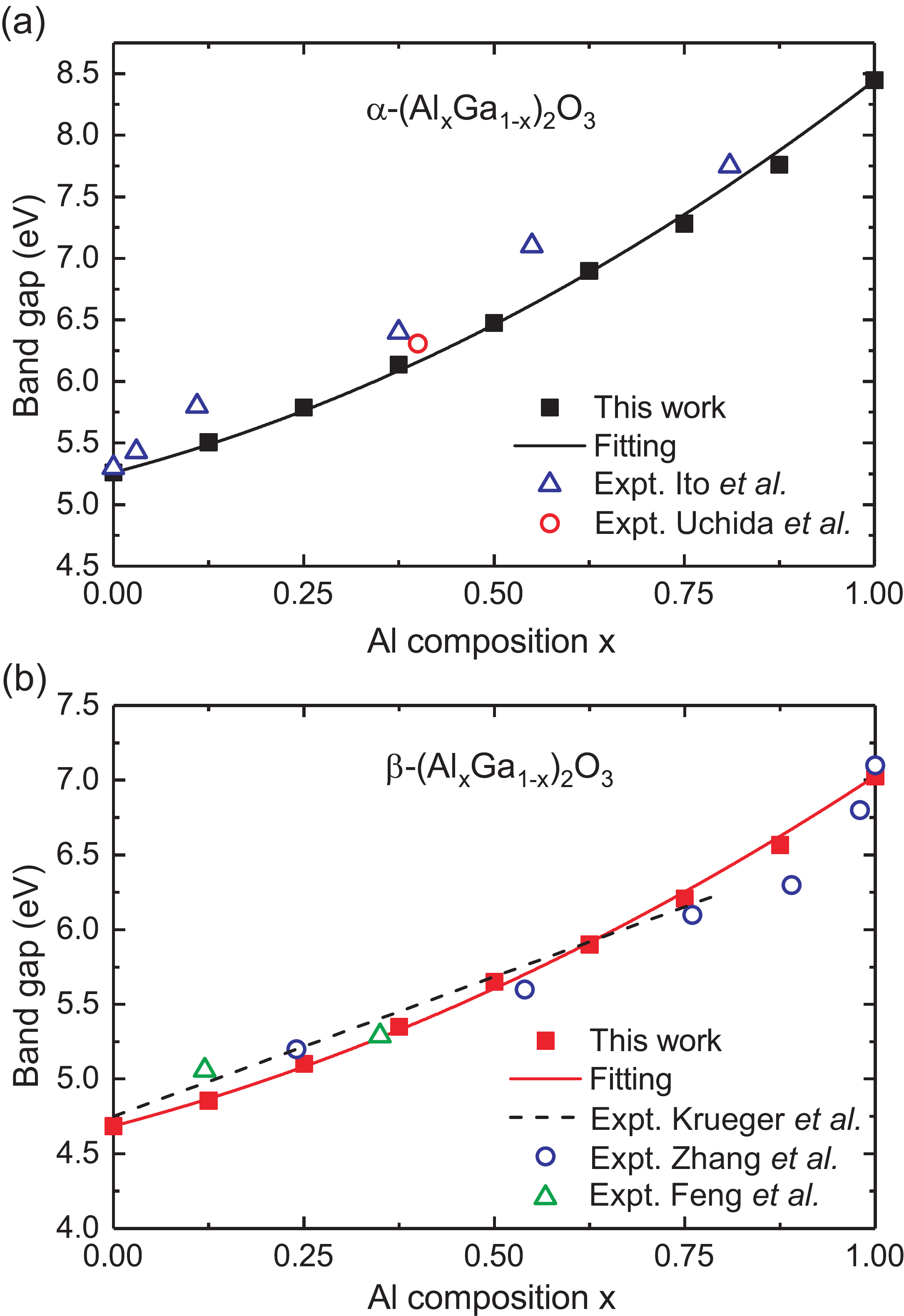}
\caption{Band gaps of (a) $\alpha$ and (b) $\beta$-(Al$_x$Ga$_{1-x}$)$_2$O$_3$ as a function of Al fraction $x$. The calculated results (solid squares) are fitted as shown as solid lines using Eq.~\ref{eq2}.  The experimental results are from the onset of photoemission inelastic losses in x-ray photoelectron spectroscopy (Ref.~\onlinecite{Krueger2016}), transmittance and x-ray photoelectron spectroscopy (Ref.~\onlinecite{Zhang2014b}), energy loss spectra (Ref.~\onlinecite{Uchida2016}), and optical transmittance and optical absorption spectra (Ref.~\onlinecite{Feng2017,Ito2012}.)
}
\label{Fig4}
\end{figure}

The calculated band gap of (Al$_x$Ga$_{1-x}$)$_2$O$_3$ alloys as a function of Al content is shown in Fig.~\ref{Fig4}.  
The band-gap bowing parameter $b$ is derived by fitting the results using:
\begin{equation} \label{eq2}
\begin{split}
E_\text{g}[(\text{Al}_{x}\text{Ga}_{1-x})_2\text{O}_3] = & (1-x) E_\text{g}[\text{Ga}_2\text{O}_3] \\
&+x E_{\text{g}}[\text{Al}_2\text{O}_3] - bx (1-x) \,.
\end{split}
\end{equation}
We obtained a bowing parameter of  1.6 eV for $\alpha$  and 1.0 eV for the $\beta$-(Al$_x$Ga$_{1-x}$)$_2$O$_3$ alloys. Overall, our results are in good agreement with the available experimental data \cite{Ito2012,Uchida2016, Krueger2016, Zhang2014b, Feng2017}, also shown in Figure~\ref{Fig4}. 
Due to the stability of the ordered AlGaO$_3$ ($x=0.5$), we can define two independent bowing parameters, one for $0\leq x\leq 0.5$ and another for $0.5\leq x \leq 1$. These are listed in the 
Supplemental Material along the bowing parameters for the VBM and CBM separately\cite{supplementalMaterial}.

\begin{figure*}
\includegraphics[width=11cm]{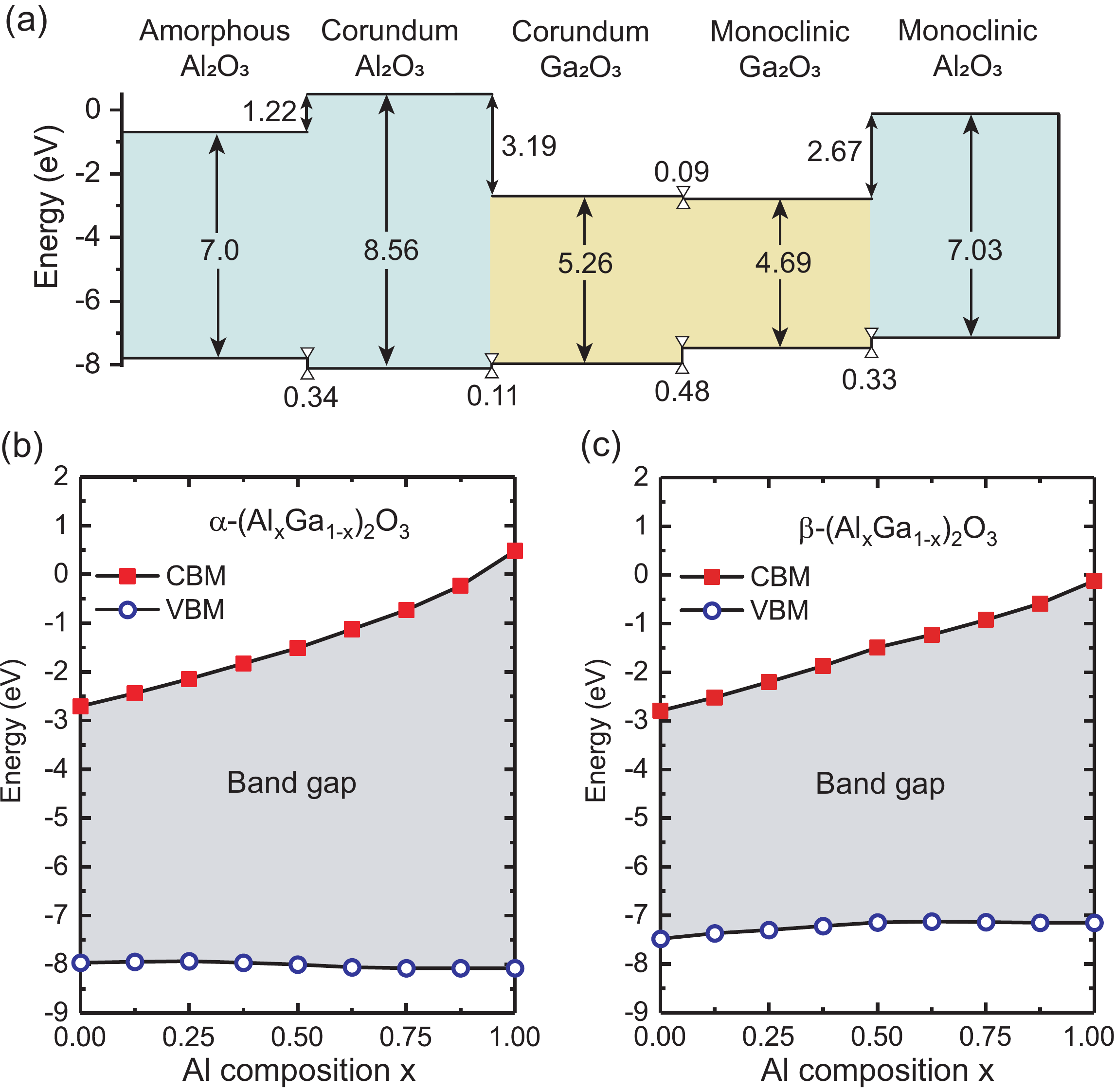}
\caption{(a) Calculated band offsets between Ga$_2$O$_3$ and Al$_2$O$_3$.  CBM and VBM positions of (c) $\alpha$ and (d) $\beta$-$(\text{Al}_{x}\text{Ga}_{1-x})_2\text{O}_3$. The zero in the energy axes corresponds to the vacuum level, determined as described in the text. }
\label{Fig5}
\end{figure*}

The band offsets between (Al$_x$Ga$_{1-x}$)$_2$O$_3$ and the parent compound Ga$_2$O$_3$ are crucial parameters in the design of electronic devices that depend on carrier confinement or on the separation between carriers and ionized impurities such as in modulation-doped field-effect transistors (MODFETs) \cite{Krishnamoorthy2017,Zhang2018}. 
We calculated the band offset between two materials using the following procedure \cite{Janotti2007b}. First, the valence-band maximum (VBM) and the conduction-band minimum (CBM) of the bulk materials are determined with respect to the averaged electrostatic potential. Then we align the averaged electrostatic potential in the two materials by performing an  an interface calculation. In this case, we used supercells composed of 12 layers of each material with two equivalent interfaces.  The supercells are constructed along non-polar directions to avoid problems resulting from the polar discontinuity at the interface and the consequent slopes in the averaged electrostatic potential in the bulk regions.
 For Ga$_2$O$_3$ and Al$_2$O$_3$ in the monoclinic phase, we chose a superlattice geometry along the [010] non-polar direction. For the corundum phase,  we constructed a superlattice  along the [100] non-polar direction. To avoid the problem of lattice mismatch, we used average in-plane lattice parameters; the out-of-plane lattice parameter is set so that each material in the superlattice has its equilibrium volume. In this way, the averaged electrostatic potentials in each side of the interface correspond the average electrostatic potential in the respective bulk materials.  This procedure, therefore, gives us the natural band offset between Ga$_2$O$_3$ and Al$_2$O$_3$.

For monoclinic Ga$_2$O$_3$ and Al$_2$O$_3$, we find a type II staggered alignment, with a valence-band offset of 0.33 eV and a conduction-band offset of 2.67 eV, as shown in Fig.~\ref{Fig5}(a). Thus, 89\% of the band offset arrises from the discontinuity in the conduction band, and only 11\% comes from the valence band. In the case of corundum Ga$_2$O$_3$ and Al$_2$O$_3$, we find a type I straddling alignment, with 0.11 eV valence-band offset and 3.19 eV conduction-band offset, also shown in Fig.~\ref{Fig5}(a), i.e., 97\% of the offset comes from the conduction band, and only 3\% from the valence band.

We also determined the absolute position of the valence and conduction bands by  taking the averaged electrostatic potential of $\alpha$ and $\beta$-Ga$_2$O$_3$ with respect to the vacuum level using surface calculations of non-polar terminations. The results are shown in the energy axis of Fig.~\ref{Fig5}(a), where we also added the band-edge positions of amorphous Al$_2$O$_3$ according to previous experimental results of band gap and valence-band offset with $\alpha$-Al$_2$O$_3$ from Ref.~\onlinecite{Filatova2015} for comparison.

Finally, we determined the band-edge positions in (Al$_x$Ga$_{1-x}$)$_2$O$_3$ with respect to that in Ga$_2$O$_3$ and Al$_2$O$_3$. The averaged electrostatic potential for a given alloy composition is derived from a linear interpolation of the averaged electrostatic potential of the constituents compounds. Figure~\ref{Fig5}(b) and (c) show the derived VBM and CBM positions for $\alpha$ and $\beta$-(Al$_x$Ga$_{1-x}$)$_2$O$_3$ as a function of Al composition. The valence-band edges only change slightly with Al concentration $x$, while most of the change occurs in the CBM. This is expected since O 2$p$ states dominate the VBM. All the band gap values, the absolute position of VBM and CBM, and the corresponding bowing parameters are listed in the Supplemental Material \cite{supplementalMaterial}.

 The results above have important implications to device design. For instance, for Al concentration of 20\% we find that the conduction-band offset between  monoclinic (Al$_{0.2}$Ga$_{0.8}$)$_2$O$_3$ and Ga$_2$O$_3$ of 0.47 eV, compared to the assumed 0.6 eV in Ref.\onlinecite{Krishnamoorthy2017} for a MODFET structure.  This relatively small conduction-band offset for 20\% Al content in the alloy can be insufficient so that electrons from the 2DEG at the  (Al$_{0.2}$Ga$_{0.8}$)$_2$O$_3$/Ga$_2$O$_3$ may stay in the Si $\delta$-doped layer in the (Al$_{0.2}$Ga$_{0.8}$)$_2$O$_3$ alloy, causing a parasitic parallel channel of conduction in the MODFET \cite{Krishnamoorthy2017}. Higher Al concentrations in the (Al$_x$Ga$_{1-x}$)$_2$O$_3$ film (Al$_x$Ga$_{1-x}$)$_2$O$_3$, resulting in increased conduction-band offsets, are required to overcome this detrimental effect.

We also note the discontinuity in the first-order derivative for the CBM and VBM at $x=0.5$ for $\beta$-(Al$_x$Ga$_{1-x}$)$_2$O$_3$. Like the discontinuities in volume and formation enthalpies, this is attributed to Al also occupying tetrahedral sites for $x>0.5$.  The kinks at $x=0.5$ in the equilibrium volume, formation enthalpies, and band-edge positions of (Al$_x$Ga$_{1-x}$)$_2$O$_3$ indicate that $\beta$-AlGaO$_3$ may well be considered an ordered compound with Al in octahedral sites and Ga in tetrahedral sites instead of a random alloy. 

In summary, we find that (Al$_x$Ga$_{1-x}$)$_2$O$_3$ alloys have much lower mixing enthalpies than (In$_x$Ga$_{1-x}$)$_2$O$_3$. The band gap of the alloys can be tuned in a wide range by changing Al composition, adding great flexibility in the design of  (Al$_x$Ga$_{1-x}$)$_2$O$_3$/Ga$_2$O$_3$-based electronic devices.  The conduction-band discontinuity comprises 89\% of the band offset between monoclinic Al$_2$O$_3$ and Ga$_2$O$_3$, and 97\% in the case of the corundum phase. Our results suggest that films with Al concentrations 
larger that 20\% are required to avoid unwanted parallel conduction channel in MODFETs based on monoclinic (Al$_x$Ga$_{1-x}$)$_2$O$_3$/Ga$_2$O$_3$ heterostructures.

\section*{Acknowledgements} 

T.W. and C.N. were supported by the II-VI Foundation, and A.J. was supported by the National Science Foundation Grant No. DMR-1652994. 
This research was supported through the use of the Extreme Science and Engineering Discovery Environment (XSEDE) supercomputer facility, National Science Foundation grant number ACI-1053575, and the Information Technologies (IT) resources at the University of Delaware, specifically the high-performance computing resources.

%

\end{document}



\title{Supplemental Material for: \\Band Gap and Band Offset of Ga$_2$O$_3$ and (Al$_x$Ga$_{1-x}$)$_2$O$_3$ Alloys}

\author{Tianshi Wang}
\affiliation{Department of Materials Science and Engineering, University of Delaware, Newark, Delaware 19716, USA}

\author{Wei Li}
\affiliation{Department of Materials Science and Engineering, University of Delaware, Newark, Delaware 19716, USA}

\author{Chaoying Ni}
   \email{cni@udel.edu}
\affiliation{Department of Materials Science and Engineering, University of Delaware, Newark, Delaware 19716, USA}

\author{Anderson Janotti}
  \email{janotti@udel.edu}
\affiliation{Department of Materials Science and Engineering, University of Delaware, Newark, Delaware 19716, USA}
%


\maketitle

We list the values of band gap $E_g$, VBM and CBM of both $\alpha$-(Al$_x$Ga$_{1-x}$) and $\beta$-(Al$_x$Ga$_{1-x}$)$_2$O$_3$ for all alloy compositions in Table~\ref{tableS1}. 
For $\beta$-(Al$_x$Ga$_{1-x}$)$_2$O$_3$, the band gap variation can be described by two different bowing parameters, $b_1$ for $0\leq x \leq 0.5$ and $b_2$ for $0.5 \leq x \leq1$. We can see this by carefully inspecting Fig.~4(b), and it can be explained by the fact that for $x=0.5$ the alloy is more stable in an ordered structure where all Ga atoms occupy the tetrahedral sites and all Al atoms the octahedral sites. 

The band-gap bowing parameters $b_1$ and  $b_2$ are determined by fitting the results to Eq.~2 where $\beta$-Ga$_2$O$_3$ and $\beta$-AlGaO$_3$ are taken as parent compounds for $0\leq x \leq 0.5$, and $\beta$-AlGaO$_3$ and $\beta$-Al$_2$O$_3$ for $0.5 \leq x \leq1$.  Similarly we define bowing parameters for VBM and CBM for the alloys.  The results are shown in Table~\ref{tableS2}, where we also list values of $b_1$ and  $b_2$ corresponding to $0\leq x \leq 0.5$ and $0.5 \leq x \leq1$.


\begin{table*}
\setlength\tabcolsep{6pt} 
\caption{\label{tableS1}
Calculated VBM and CBM positions with respect to the to vacuum level, and direct band gaps of  $\alpha$- and $\beta$-(Al$_x$Ga$_{1-x}$)$_2$O$_3$.   }
\begin{ruledtabular}
\begin{tabular}{lccccccccc}
\rule{0pt}{10pt} & \multicolumn {9}{c}{ $\alpha$-(Al$_x$Ga$_{1-x}$)$_2$O$_3$}\\
\rule{0pt}{10pt}Al content $x$ & 0& 0.125& 0.25 & 0.375 & 0.5 & 0.625 & 0.75 & 0.875&1 \\
\hline
CBM (eV) & -2.71 &-2.45 &-2.15 &-1.83 &-1.52 &-1.13 &-0.74 &-0.24 &0.48 	\\
VBM (eV)&-7.97 &	-7.95 &	-7.94 &	-7.97 &	-8.01 &	-8.07 &	-8.08 &	-8.08 &	-8.08 \\
$E_\textrm{g}$ (eV) &5.26 &	5.50 &	5.78 &	6.13 &	6.49 &	6.93 &	7.34 &	7.84 &	8.56  \\
\hline
\rule{0pt}{10pt} & \multicolumn {9}{c}{ $\beta$-(Al$_x$Ga$_{1-x}$)$_2$O$_3$}\\
\rule{0pt}{10pt}Al content $x$ & 0& 0.125& 0.25 & 0.375 & 0.5 & 0.625 & 0.75 & 0.875&1 \\
\hline
CBM (eV) & -2.80 &	-2.52 &	-2.21 &	-1.87 &	-1.50 &	-1.23 &	-0.93 &	-0.59 &	-0.13 	\\
VBM (eV)& -7.49 &	-7.38 &	-7.31 &	-7.22 &	-7.15 &	-7.13 &	-7.14 &	-7.16 &	-7.16 \\
$E_\textrm{g}$ (eV) & 4.69 &	4.86 &	5.1 &	5.35 &	5.65 &	5.9 &	6.21 &	6.57 &	7.03  \\
\end{tabular} \\
\end{ruledtabular}
\end{table*}

\begin{table*}
\setlength\tabcolsep{6pt} 
\caption{\label{tableS2}
Bowing parameter $b$ for band gap and band edge positions.   }
\begin{ruledtabular}
\begin{tabular}{ccccc}
\rule{0pt}{10pt}Material  & Al content $x$ range & $b^{E_\textrm{g}}$ (eV)  & $b^{CBM}$ (eV) & $b^{VBM}$ (eV)  \\
\hline

              & $0\leq x\leq 0.5$ & 0.32 &0.14 &-0.18  	\\
 $\alpha$-(Al$_x$Ga$_{1-x}$)$_2$O$_3$   & $0.5\leq x\leq 1$ &	0.74 &	0.88  &0.15 \\
              & $0\leq x\leq 1$ &	1.6 &	1.67 &	 -0.05  \\
\hline

  &$0\leq x\leq 0.5$ &	0.32 &0.24 &	-0.08 	\\
 $\beta$-(Al$_x$Ga$_{1-x}$)$_2$O$_3$ &$0.5\leq x\leq 1$ &	0.54 &	0.48 &	-0.06  \\
  & $0\leq x\leq 1$ &	1.0 &	0.40&	-0.61   \\
\end{tabular} \\
\end{ruledtabular}
\end{table*}